\documentclass[preprint,12pt]{elsarticle}

\usepackage{amsthm}
\usepackage{amsmath}
\usepackage{float}
\usepackage{url}
\usepackage{yfonts}
\usepackage{comment}
\usepackage{graphicx}
\usepackage{indentfirst}
\usepackage{amsfonts}
\usepackage{latexsym,amsmath,epsfig,amssymb,graphics}

\usepackage[ruled,vlined,linesnumbered]{algorithm2e}
\def \froot {{\tt logcf}}
\def \MM {{\tt Mathematica}}
\def \inte {{\tt RootIntervals}}
\def \cf {{\tt CF}}
\def \sle {{\tt Sleeve}}
\def \eign {{\tt eigensolve}}

\def \ZZ {{\mathbb Z}}

\def \RR {{\mathbb R}}

\def \algcf{{\tt cf}}
\def \sign{{\rm sign}}
\def \abs {{\rm abs}}
\def \gcd {{\rm gcd}}
\def \deg {{\rm deg}}
\def \algm{{\tt main}}
\def \dec {{\tt dec}}
\def \up {{\tt logup}}

\def \less {{\tt lessOne}}
\def \lb {{\tt loglb}}

\newtheorem{thm}{Theorem}

\newdefinition{rmk}{Remark}
\newproof{pf}{Proof}
\newproof{pot}{Proof of Theorem \ref{thm2}}

 \newdefinition{note}{Notation}

 \newdefinition{definition}{Definition}

 \newtheorem{cor}[thm]{Corollary}

\usepackage{amssymb}

\journal{Computers $\&$ Mathematics with Applications}

\begin{document}

\begin{frontmatter}



\title{\froot: An Efficient Tool for Real Root Isolation}


\author{Liyun Dai\corref{cor}}
\ead{dailiyun@pku.edu.cn}
\author{Bican Xia}
\ead{xbc@math.pku.edu.cn}
\cortext[cor]{Corresponging author}

\address{LMAM $\&$ School of Mathematical Sciences, Peking University}

\begin{abstract}

  This paper revisits an   algorithm for isolating real roots of  univariate
  polynomials based on continued fractions. It follows  the work of Vincent,	Uspensky, Collins and Akritas, Johnson and Krandick.
  We use  some tricks, especially  a new algorithm  for computing an upper bound of positive roots.  In this way, the algorithm of   isolating real roots
  is improved.   The complexity of our method for computing an upper bound of positive roots is  $O(n\log(u+1))$ where $u$ is the optimal upper bound satisfying
  Theorem \ref{thm:log} and  $n$ is the degree of the polynomial. Our  method has been implemented as a  software package \froot\ using \texttt{C++} language. For many benchmarks \froot \  is  two or three times
  faster   than  the function {\tt RootIntervals} of \MM. And it  is much faster than   another continued fractions based software \cf, which seems to be one of the
  fastest available open software for exact real root isolation. For those  benchmarks which  have only real roots, \froot\ is
  much faster than \sle\ and \eign\ which are based on numerical computation.
\end{abstract}


\end{frontmatter}


\section{Introduction }
\label{}
Real root isolation of univariate polynomials with integer coefficients is one of the fundamental tasks in computer algebra as well as in many applications ranging from computational geometry to quantifier elimination. The problem can be stated as: given a polynomial $P\in \ZZ[x]$, compute for each of its real roots an interval with rational endpoints containing it and being disjoint from the intervals
computed for the other roots.  The methods of isolating real root can be divided into three kinds.  The first kind consists of the subdivision algorithms using counting techniques based  on, {\it e.g.}, the Strum theorem or
Descartes' rule of signs.  This kind of methods count the sign changes (of Sturm sequence or coefficients of some polynomials) in the considered interval and if the sign changes reach $1$ or $0$, the procedure returns from this interval.
Otherwise it subdivides the interval and compute recursively. The symbolic implementations of these methods can be found in \cite{collin76,rou04} and the symbolic-numberic algorithms implementations can be found in \cite{rou04,eig05,eig08,meh11}. 

The second kind takes use of the continued fraction (CF) algorithms \cite{akr08,tsi08,sha08}. These methods are highly efficient and competitive \cite{rou04,akr05,hemmer09}. Especially,  \cite{hemmer09} provides a test datasets   consisting of 5000 polynomials from many
different settings,  with results indicating that there is no best method overall. However one can say that for most instances the solvers based on CF are among
the best methods. In this paper we modify a real root isolation algorithm based on CF method to obtain a more efficient tool \froot.

The third kind is based on Newton-Raphson method and interval arithmetic.
The search space is subdivided until it contains only a single real root and Newton's method converges. When the polynomial is sparse and has very high degree, this method will be much faster than other methods. The symbolic implementations of this kind of methods can be found in \cite{xia06,xia07} and the numeric implementations
can be found in \cite{kla93,rump99}.

Those methods based on CF compute the continued fraction expansion of the real roots of a polynomial in order to compute isolating intervals for real roots. One important step
is the computation of upper bounds of the positive real roots of some polynomials. There are  several classic methods to compute such upper bounds, such as Cauchy bounds, Lagrange-MacLaurin  bounds and Kioustelidis' bounds. There are many recent works about the upper bound of the positive roots of univariate polynomials \cite{hong98,ste05,akr05,akr06,akr08}. Some methods for computing such bounds are of $O(n)$ complexity but the results are very coarse like Cauchy bounds. Some methods are of $O(n^2)$ complexity but their bounds are sharper such as the method presented in \cite{akr08}. The balance between precision and effect for computing such upper bounds has to be taken into account.

We provide a new method for computing such bounds with time complexity $O(n\log(u+1))$, where $u$ is the optimal upper bound satisfying Theorem \ref{thm:log}. Besides, compared
with \cite{akr08}, when  Algorithm \ref{alg:less} return true (the upper bound is less than $1$), our upper bound is at most two times that in \cite{akr08}. In this way, the algorithm of   isolating real roots is improved.  Our  method has been implemented as a  software package \froot\ using \texttt{C++} language. For many benchmarks \froot \  is  two or three times
faster   than  the function {\tt RootIntervals} of \MM. And it  is much faster than   another continued fractions based software \cf, which seems to be one of the
fastest available open software for exact real root isolation. For those  benchmarks which  have only real roots, \froot\ is much faster than \sle\ and \eign\ which are based on numerical computation.

The rest of this paper is organized as follows. Section 2 reviews the
main algorithm for real root isolation based on CF. Section 3 presents a new algorithm for
computing an upper bound of positive roots. Section 4 lists some tricks used in \froot.  Section 5 lists the comparative experimental results of our algorithm and other software. 

\section{Algorithm based on CF}
In this section, we first recall Descartes' rule of signs, which gives a bound on the number of positive real roots. Then the Vincent theorem, which can ensure
the termination of algorithms based on CF, is presented. Finally, we review an algorithm of real root isolation based on CF.

As usual, $\deg(p)$ denotes the degree of univariate polynomial $p$. The derivative of polynomial $p$ with respect to the only variable is denoted by $p'$ and $\gcd(f,g)$ means the greatest common divisor of polynomials $f$ and $g$.

\begin{note}[Sign variation]
Let $S=\left\{ a_0,a_1,\ldots,a_n \right\}$ be a finite sequence of non-zero real numbers. Define $V(S)$, the {\em sign variation} of $S$, as follows.
\[V(S)=0\ \text{ if } |S|\le1,\]
\[  V(a_0,\ldots,a_{n-1},a_n)=  \left\{\begin{aligned}
 &  V(a_0,\ldots,a_{n-1})+1 \text{ if }a_{n-1}a_n<0;\\
&V(a_0,\ldots,a_{n-1}), \text{ otherwise}.\\
	\end{aligned}
	\right.
\]
If some elements of $S$ are zero, remove those zero-elements to get a new sequence and define $V(S)$ to be the sign variation of this new sequence.
\end{note}



\begin{thm}[Descartes' rule of signs] \label{thm:des}
  Suppose $p=\sum_{i=0}^na_ix^i\in\RR[x]$ has $m$ positive real roots, counted with multiplicity. Set $V(p)=V(a_0,a_1,\ldots,a_n)$. Then $m\le V(p)$, and $V(p)-m$ is even.
\end{thm}

\begin{thm}[Vincent's theorem]\label{thm:vin}
  Let $P(x)$ be a real polynomial of degree $n$ which has only simple roots. It is possible to determine a positive quantity $\delta$ so that for every pair of positive real numbers $a$ and $b$ with $|b-a| < \delta$, the coefficients sequence of every transformed polynomial of the form
  $  P(x) = (1+x)^{n}P(\frac{a+bx}{1+x}) $
		  has exactly 0 or 1 sign variation. The second case is possible if and only if $P(x)$ has a single root within $(a,b)$.
\end{thm}

\begin{algorithm}\label{alg:main}
\SetAlgoCaptionSeparator{.}
\caption{\algm}
\DontPrintSemicolon
\KwIn{ A non-zero polynomial $P(x)\in \ZZ[x] $. }
\KwOut{$I$, a set of real root isolating intervals of $P(x)$. }
$I=\emptyset$; \;
\If {$\deg(P)= = 0$} {return $I$;}
$P=\frac{P}{\gcd(P,P')}$; \tcc{ square free}
\If {$P(0)= = 0$ } { $I.add([0,0])$; \tcc{ add $[0,0]$ to set $I$}
\dec($P$); \tcc {Algorithm \ref{alg:dec}} }
$I$.addAll(\algcf($P$)); \;
\tcc { add all the positive root intervals to set $I$ } \tcc{ \algcf\ is described as Algorithm \ref{alg:cf}}
$p=-p$;\;
$I$.addAll(\algcf($P$)); 
\end{algorithm}

CF based procedures will continue subdividing the considered interval into two subintervals and make a one to one map from $(a,b)$ to $(0,+\infty)$ by $  P(x) = (1+x)^{n}P(\frac{a+bx}{1+x})$ until $V(P)$ equals $1$ or $0$. Therefore, Theorem \ref{thm:vin} guarantees the termination of these  procedures.

\begin{definition} As in \cite{akr05}, we define the following transformations for a univariate polynomial $P(x)$.
  \begin{eqnarray*}
  R(P(x))=x^n(P(\frac{1}{x})),\\
  H_\lambda(P(x))=P(\lambda x),\\
  T(P(x))=P(x+1).
  \end{eqnarray*}
\end{definition}

$T(P)$  is also called	Taylor shift one \cite{ger04,joh05}. In our experiments when Algorithm \ref{alg:up} is used for computing upper bounds, $T(P)$  takes  more than
ninety percent of running time\footnote{the result of  GNU gprof.}.  We have considered methods in \cite{ger04} for computing $T(P)$, but   finally we chose the  classical
method (Horner's method) for its simplicity. In future work we will use Divide \& Conquer method which is the fastest in \cite{ger04}. We think this substituting
will still improve the performance of our method.

\begin{algorithm}\label{alg:dec}
\SetAlgoCaptionSeparator{.}
\caption{\dec}
\DontPrintSemicolon
\KwIn{ $P=a_nx^n+a_{n-1}x^{n-1}+\cdots+a_1x+a_0,n>0 $. }
\KwOut{$P=a_nx^{n-1}+a_{n-1}x^{n-2}+\cdots+a_2x+a_1$ . }
\end{algorithm}

\begin{algorithm}\label{alg:lb}
\SetAlgoCaptionSeparator{.}
\caption{\lb}
\DontPrintSemicolon
\KwIn{ $P\in\ZZ[x] $. }
\KwOut{$root\_lb$, a lower bound of positive roots of $P$. }
$P=R(P)$;\;
$root\_lb=$\up($P$); \tcc{\up\ is described as Algorithm \ref{alg:up}}
\end{algorithm}

\begin{definition}
$  intvl(a,b,c,d)=  \left\{\begin{aligned}
&  (\min\left\{ \frac{a}{c},\frac{b}{d} \right\},\max\left\{ \frac{a}{c},\frac{b}{d} \right\} ) &\text{ if } cd\neq0;\\
& (0,\infty), &\text{ otherwise}.\\
	\end{aligned}
	\right.
$
\end{definition}

Using the above notations and definitions, an algorithm for isolating all the real roots of a nonzero univariate polynomial is described as Algorithm \ref{alg:main}.
Algorithm \ref{alg:cf}, which has only a little modification of the algorithm in \cite{akr08}, is presented here to make our subsequent description clearer.

\section{A new algorithm of computing upper bounds}

One key ingredient of CF based methods is the computation of upper bounds of the positive real roots of some polynomials. We give in Theorem \ref{thm:log} a new characteristic of such upper bounds of univariate polynomials. A new algorithm based on this theorem, Algorithm \ref{alg:up}, is proposed for computing upper bounds of positive real roots.

\begin{algorithm}\label{alg:cf}
\SetAlgoCaptionSeparator{.}
\caption{\algcf}
\DontPrintSemicolon
\KwIn{ A squarefree polynomial $F \in \ZZ[x] \setminus \{0\}$. }
\KwOut{ $roots$, a list of isolating intervals of positive roots of $F$. }

 $roots=\emptyset$;
  $s=V(F)$;\;

$intstack=\emptyset$;
$intstack$.add($\{1,0,0,1,F,s\}$);\;

\While{$intstack\neq \emptyset$} {
  $\{a,b,c,d,P,s\}=intstack.$pop();\tcc{pop  the first element}
  $\alpha=\lb(P)$;\;
  \If{$\alpha\ge1$ }{
	$\{ a,c,P \}=\{ \alpha  a,\alpha c ,H_\alpha(P) \}$;
	$\{ b,d,P \}=\{   a+b,c+d ,T(P) \}$;

	\If { $P(0)= = 0$}{
	  $roots$.add$([\frac{b}{d},\frac{b}{d}])$ ;
	  $P=\frac{P}{x}$;\;
	}
	$s=V(P)$;\;
	\If {$s= = 0$}{
	  continue;\;
	}
	\ElseIf{$s= = 1$}{
			
	  $roots$.add($intvl(a,b,c,d)$);
	  continue;\;
	}

  }
  $ \left\{ P_1,a_1,b_1,c_1,d_1,r \right\}=\left\{ T(P),a,a+b,c,c+d,0 \right\}$

\If{$P_1(0)= =0$}{
  $roots$.add($[\frac{b_1}{d_1},\frac{b_1}{d_1}]$);
  $P_1=\frac{P_1}{x};r=1$;\;
}
$s_1=V(P_1)$;
$\left\{ s_2,a_2,b_2,c_2,d_2 \right\}=\left\{ s-s_1-r,b,a+b,d,c+d \right\}$;

\If {$s_2>1$ }{
  $ P_2= (x+1)^{\deg(P)}T(P)$;\;
  \If {$P_2(0)= = 0$}{
	$P_2=\frac{P_2}{x}$;
	$s_2=V(P_2)$;\;
  }
}
\If{$s_1= = 1$ }{

	  $roots$.add($intvl(a_1,b_1,c_1,d_1)$);\;
}
\ElseIf{$s_1>1$ }{

$intstack$.add($\{a_1,b_1,c_1,d_1,P_1,s_1\}$);\;
}

\If{$s_2= = 1$}{

	  $roots$.add($intvl(a_2,b_2,c_2,d_2)$);\;
}
\ElseIf{$s_2>1$ }{

$intstack$.add($\{a_2,b_2,c_2,d_2,P_2,s_2\}$);\;
}
}
\end{algorithm}

\begin{thm} \label{thm:log}
  Suppose   $P=a_nx^n+a_{n-1}x^{n-1}+\cdots+a_1x+a_0\ (a_n>0)$  is a univariate polynomial in $x$ with real coefficients.  Then  a nonnegative number $u$ is an upper bound of positive roots of $P$ if $u$   satisfies $\min_{j=0}^{n}\left\{  \sum_{i=j}^n a_i u^{i-j}\right\}\ge0$.
\end{thm}
\begin{pf}
  If $n= =0$, then $P$ is a nonzero constant and any positive number is its upper bound of positive roots.

  Otherwise, if $b>u$,  we claim that $\sum_{i=j}^na_ib^{i-j}> \sum_{i=j}^na_iu^{i-j}$ for any $j= 0,\ldots,n-1$.

  When $j=n-1$, $\sum_{i=n-1}^na_ib^{i-n+1}-\sum_{i=n-1}^na_iu^{i-n+1}=a_n(b-u)>0.$ The claim holds.

  Assume the claim holds  when $j=k$. When $j=k-1$,  $\sum_{i=k-1}^na_ib^{i-k+1}=\left(\sum_{i=k}^na_ib^{i-k}\right)b+a_{k-1} $. By assumption
  $\sum_{i=k}^na_ib^{i-k}>\sum_{i=k}^na_iu^{i-k}\ge0$. Since $b>u\ge0$, $\left(\sum_{i=k}^na_ib^{i-k}\right)b>\left (\sum_{i=k}^na_iu^{i-k} \right)u  $
  and $\sum_{i=k-1}^na_ib^{i-k+1}> \sum_{i=k-1}^na_iu^{i-k+1}$. So  $\sum_{i=j}^na_ib^{i-j}> \sum_{i=j}^na_iu^{i-j}$ for any $j= 0,\ldots,n-1$.

  By the above claim,   $P(b)=\sum_{i=0}^na_ib^i>0$ when  $b>u$. Because $b$ is arbitrarily chosen, $u$ is an upper bound of the positive roots of $P$.

\end{pf}

The following theorem was given by Akritas et al. in  \cite{akr06,akr08}, which computes positive root upper bounds of univariate polynomials.

\begin{thm}[Akritas-Strzebo\'{n}ski-Vigklas, \cite{akr06}] \label{thm:qup}

  Let $P(x)=a_nx^n+a_{n-1}x^{n-1}+\cdots+a_0\ (a_n>0)$ be a polynomial with real coefficients and let $d(P)$ and $t(P)$ denote the degree and the number of its terms, respectively.

  Moreover, assume that $P(x)$ can be written as\begin{equation}
	P(x)=q_1(x)-q_2(x)+q_3(x)-q_4(x)+\cdots +q_{2m-1}(x)-q_{2m}(x)+g(x)
	\label{eq:1}
  \end{equation}
  where all the coefficients of polynomials $q_i(x)$ $(i=1,2,\ldots,2m)$ and $g(x)$ are positive. In addition,
  assume that for $i=1,2,\ldots,m$ we have
  \begin{equation*}
	q_{2i-1}(x)=c_{2i-1,1}x^{e_{2i-1,1}}+\cdots+c_{2i-1,t_{2i-1}}x^{e_{2i-1,t_{2i-1}}}
  \end{equation*}
  and
  \begin{equation*}
	q_{2i}(x)=b_{2i,1}x^{e_{2i,1}}+\cdots+b_{2i,t_{2i}}x^{e_{2i,t_{2i}}}
  \end{equation*}
  where $e_{2i-1,1}=d(q_{2i-1})$, $e_{2i,1}=d(q_{2i})$, $t_{2i-1}=t(q_{2i-1}),$ and $t_{2i}=t(q_{2i})$   and the exponent of each term in $q_{2i-1}(x)$ is greater than the exponent of each term in $q_{2i}(x)$. If for all indices $i=1,2,\ldots,m$, we have
  \begin{equation*}
	t(q_{2i-1})\ge t(q_{2i}),
  \end{equation*}
  then an upper bound of the values of the positive roots of $p(x)$ is given by

  \begin{equation}\label{eq:2}
	up= \max_{i=1,2,\ldots,m}\left\{ \max_{j=1,2,\ldots,t_{2i}}\left\{ \left( \frac{b_{2i,j}}{c_{2i-1,j}} \right)^{\frac{1}{e_{2i-1,j}- e_{2i,j} } } \right\}  \right\}
  \end{equation}
  for any permutation of the positive coefficients $c_{2i-1,j}, $ $j=1,2,\ldots,t_{2i-1}$.
  Otherwise, for each of the indices $i$ for which we have
  \begin{equation*}
	t_{2i-1}<t_{2i},
  \end{equation*}
  we {\bf break up} one of the coefficients of $q_{2i-1}(x)$ into $t_{2i}-t_{2i-1} +1$ parts, so that now $t(q_{2i} ) = t(q_{2i-1})$ and apply the same formula (\ref{eq:2}) given above.

\end{thm}


We shall show in Theorem \ref{thm:com1} that the bound given by Theorem \ref{thm:log} is better than that given by Theorem \ref{thm:qup}.

\begin{thm}\label{thm:com}
   Let $P(x)=a_nx^n+a_{n-1}x^{n-1}+\cdots+a_0\ (a_n>0)$ be a polynomial with real coefficients and   $u$ denote an upper bound of positive roots of $p$ obtained by Theorem \ref{thm:qup}, then $\min_{k=0}^{n}\left\{  \sum_{i=k}^n a_i u^{i-k}\right\}\ge0$.

\end{thm}

\begin{pf}

  For every $ a_i<0$, by Theorem \ref{thm:qup}, there exist  $c_{i_1}x^{e_{i_1}}$ and $b_{i_2}x^{e_{i_2}},$  respectively, such that
    $e_{i_1}>e_{i_2}$ and $c_{i_1}u^{e_{i_1}}\ge b_{i_2}u^{e_{i_2}}$. By  Theorem \ref{thm:qup} $b_{i_2}x^{e_{i_2} }$ is the term $-a_ix^i$
  and $c_{i_1}x^{e_{i_1} }$ is either a whole or a part (broken up by Theorem \ref{thm:qup}) of a positive term of $p$.

  For every  $a_j>0$, by Theorem \ref{thm:qup}, $\left( \sum_{a_i<0,e_{i_1}=j }c_{i_1} \right)\le a_{j}$. So $\sum_{i=k}^na_iu^i\ge \sum_{i=k,a_i<0}^n \left( c_{i_1}u^{e_{i_1}}-
  b_{i_2}u^{e_{i_2}} \right)\ge 0 $ for any $k= 0,1,\ldots,n$. Then $\sum_{i=k}^n a_i u^{i-k}\ge0 $ for any  $k= 0,1,\ldots,n$ and
  $\min_{k=0}^{n}\left\{  \sum_{i=k}^n a_i u^{i-k}\right\}\ge0$.

\end{pf}


\begin{thm}\label{thm:com1}

  Let $P(x)=a_nx^n+a_{n-1}x^{n-1}+\cdots+a_0\ (a_n>0)$ be a polynomial with real coefficients. Let  $u_1$ denote the optimal upper bound of
  positive real roots satisfying Theorem \ref{thm:log} and $u_2$ denote the optimal upper bound of positive real roots satisfying Theorem \ref{thm:qup}, then $u_1\le u_2$ and the  strict inequality can hold.
\end{thm}
\begin{pf}
  By Theorem \ref{thm:com}, $u_1\le u_2$.

  Let $P(x)=x^2+x-2$. Then $u_2=\sqrt{2}$ and $u_1=1$. So
  $u_1<u_2$ for this $P$.
\end{pf}

\begin{thm}
  Let $P(x)=a_nx^n+a_{n-1}x^{n-1}+\cdots+a_0\ (V(P)> 0)$ be a polynomial with real coefficients. Let  $u$ denote the output of Algorithm \ref{alg:up} and $u_1$ denote the optimal upper bound of $P$ satisfying
 Theorem \ref{thm:log}. When $u$ is less than or equal to $1$, $u<2u_1$.

\end{thm}

\begin{algorithm}\label{alg:less}
\SetAlgoCaptionSeparator{.}
\caption{\less}
\DontPrintSemicolon
\KwIn{ $P=a_nx^n+a_{n-1}x^{n-1}+\cdots+a_1x+a_0\in\ZZ[x], \exists a_i,a_na_i<0 $. }
\KwOut{ true: the positive root bound of $P$ must be less than $1$;\\\ \ \ \ \ \ \ \ \ \ \ \ \ \    false: cannot determine whether the bound is less than $1$. }

$start=n-1$;\;
$lastNeg=0$;\;
$hSign=\sign(a_n)$;\;

\While{$\sign(a_{lastNeg})+hSign\neq 0$ }{
  $lastNeg=lastNeg+1$;\;

}

\While{$\sign(a_{start})+hSign\neq 0$ }{
  $start=start-1$;\;

}
$cfSum=\abs(a_n)$;\;
$i=n-1$;\;
$j=start$;\;
$last=start$;\;
\While{$i\ge lastNeg-1 \text{ and } j\ge lastNeg-1$}
{
	\If{$\sign(cfSum)<0$}
	{
	  \While{$i>last\text{ and } \sign(a_i)\neq hSign$ }{
		$i=i-1$;
	  }

	 \If{$i= = last$}{
	   \Return {false;}
	 }
	 $cfSum=cfSum+\abs(a_i)$;\;
	 $i=i-1$;\;

	}\Else{
	  \If{$j= = lastNeg-1$ }{
		\Return{ true;}

	  }
	  \While{$j\ge lastNeg \text{ and } \sign(a_j)+hSign\neq 0$ }{
		$j=j-1$;
	  }
	  $cfSum=cfSum-\abs(a_j)$;\;
	  $last=j$;\;

	}

}

\Return {true; }

\end{algorithm}

\begin{algorithm}[H]\label{alg:up}
\SetAlgoCaptionSeparator{.}
\caption{\up}
\DontPrintSemicolon
\KwIn{$P=a_nx^n+a_{n-1}x^{n-1}+\cdots+a_1x+a_0\in\ZZ[x], \exists a_i,a_na_i<0. $ }
\KwOut{ an upper bound of the positive roots of $P$. }
$start=n-1$;
$lastNeg=0$;
$hSign=\sign(a_n)$;
$base=1$;\;

\If{$\neg$\less($P$) }{
  return $2$;
}
\While{$\sign(a_{lastNeg})+hSign\neq 0$ }{
  $lastNeg=lastNeg+1$;\;

}

\While{$\sign(a_{start})+hSign\neq 0$ }{
  $start=start-1$;\;

}

$i=n$;\;
\While{$i= = n$}{

  $i=n-1$;\;
  $j=start$;\;
  $cfSum=\abs(a_n)$;\;

	\While{$i\ge lastNeg-1 \text{ and } j\ge lastNeg-1$}{
		\If{$\sign(cfSum)<0$}
		{
		  \While{$i>j\text{ and } \sign(a_i)\neq hSign$ }{
			$i=i-1$;
		 }
		 \If {$i= = j$}{
		   break;
		 }
		 $cfSum=cfSum+\abs(a_i)2^{(n-i)base}$;\;
		 $i=i-1$;\;

	   }\Else{

		 \If{$j= = lastNeg-1$ }{
			$j=lastNeg-2$;
			break;\;

		 }

		\While{$j\ge lastNeg \text{ and } \sign(a_j)+hSign\neq 0$ }{
			$j=j-1$;
		}

		 $cfSum=cfSum-\abs(a_i)2^{(n-j)base}$;\;

		$j=j-1$;\;

	   }

  }
  \If{$j= = lastNeg-2$}{
	$base=base+1$;
	$i=n$;\;

  }

}
\Return{$\frac{1}{2^{base-1}}$; }

\end{algorithm}

\begin{pf}
  In Algorithm \ref{alg:up}, if  $\frac{1}{2^{base}}\ge u_1,$ then $\min_{j=0}^{n}\left\{ \sum_{i=j}^na_i\left( {\frac{1}{2^{base}} }
  \right)^{i-j}\right \}\ge 0$ by the proof of Theorem \ref{thm:log} and thus 
  the loop does not terminate at this step.
  So when Algorithm \ref{alg:up} returns, $base$ must satisfy $\frac{1}{2^{base}}<u_1$. Therefore, the output $u=\frac{1}{2^{base-1}}$ and $u<2u_1$.
  Obviously, this algorithm will terminate.

  Furthermore,  $\min_{j=0}^{n-1}\left\{ \sum_{i=j}^na_i\left( {\frac{1}{2^{base-1}} } \right)^{i-j}\right \}\ge 0$ by Theorem \ref{thm:log}. So, $u=\frac{1}{2^{base-1}}$ is an upper bound of $p$.
\end{pf}

\begin{cor}
  Let $P(x)=a_nx^n+a_{n-1}x^{n-1}+\cdots+a_0$ $ (V(P)> 0)$ be a polynomial with real coefficients. Set $u$ to be the optimal upper bound of positive roots of $P$ satisfying Theorem
  \ref{thm:log}. Then  Algorithm \ref{alg:up} costs at most $O(n\log(u+1))$  additions and multiplications.
\end{cor}

\section{Tricks}
{\bf Variable substitution}
If $P(x)\in \ZZ[x]$ and $P(x)=P_1(x^k)\ (k>1),$ then substitute $y=x^k$ in $P$. Obviously, $\deg(P_1,y)=\frac{\deg(P,x)}{k}$. We first isolate the real roots of $P_1$ then
obtain the real roots of $P$. We can see in Figure \ref{fig:2} that degree is a key fact affecting the running time. Using this trick, we can greatly reduce the running time of ${\it ChebyshevT}$
and {\it ChebyshevU} when each term of the polynomials is of even degree. The running time on such polynomials can be found in Table \ref{tab:mm2}. The same trick was also taken into account in \cite{johnson06}.

{\bf Incomplete termination check}
If $P(x)\in \ZZ[x]$ and $V(P)= 2$, we may try to check whether the sign of $P(1)$ is the same as the sign of the leading coefficient of $P$. If they are not the same, then $P$  has one positive root in $(0,1)$ and the other one in $(1,+\infty)$. So, we can terminate this subtree. Since the whole \froot\ procedure is a tree and \froot\ spends more than 90 percent of the
total time on computing $T(P)$, this trick may improve the efficiency of the algorithm greatly.

\section{Experiments }
\subsection{Implementation}
The main algorithm for isolating real roots based on our improvements has been implemented as a \texttt{C++} program, \froot \footnote{The program can be downloaded through \url{http://www.is.pku.edu.cn/~dlyun/logcf}}. Compilation was done using g++ version 4.6.3 with optimization flags -O2.
We use {\tt Singular} \cite{singular} to read polynomials from files or standard input and to eliminate multi-factors of polynomials. We use the GMP\footnote{ \url{http://gmplib.org/}}
(version 5.05), arbitrary-length integers libraries, to deal with big integer computation.
All the benchmarks listed were computed on a 64-bit Intel(R) Core(TM) i5 CPU 650 @ 3.20GHz with 4GB RAM memory and Ubuntu 12.04 GNU/Linux.

\subsection{Benchmarks }
 \subsubsection{$W_n$}
 {\it Wilkinson} polynomials: $W_n=\Pi_{i=1}^n(x-i)$. The integers $1,2,\ldots,n $ are  all the real roots of $W_n$.
  \subsubsection{$mW_n$}
  Modified {\it Wilkinson} polynomials: $mW_n=W_n-1$.

  If $n>10$, $mW_n$ has $n$ simple real roots but most of them are irrational.
 \subsubsection{$IW_n$}
 The distance between  $W_n$'s two  nearest real roots  is  $1$ and the distance between $mW_n$'s two nearest real roots  is nearly $1$. 
 We construct new polynomials $IW_n=\Pi_{i=1}^n(ix-1)$, which have a completely different distance between any two nearest real roots.
 \subsubsection{$ mIW_n$ }
 We modify $IW_n$  into $mIW_n=IW_n-1$ for the same purpose  as we construct $mW_n$. Most real roots of $mIM_n$  become irrational.
 \subsubsection{$T_n$} {\it ChebyshevT} polynomials: $T_0=1,T_1=x,T_{n+1}=2xT_n-T_{n-1}$. $T_n$ has $n$ simple real roots.
 \subsubsection{$U_n$} {\it ChebyshevU} polynomials: $U_0=1,U_1=2x,U_{n+1}=2xU_n-U_{n-1}$. $U_n$ has $n$ simple real roots.
 \subsubsection{$L_n$}
 {\it Laguerre}  polynomials: $L_0=1$,$L_1=1-x$,$L_{n+1}(x)=\frac{  (2n+1-x )L_n(x)-  nL_{n-1 }(x)}{(n+1) }$. 
Obviously, $n!L_n$ is a polynomial with integer coefficients.
 \subsubsection{$M_n$} {\it Mignotte} polynomials: $x^n-2(5x-1)^2$. If $n$ is odd, $M_n$ has three simple real roots. If $n$ is even, it has four simple real roots.
 \subsubsection{$R(n,b,r) $} Randomly generated polynomials: $R(n,b,r)$=$a_nx^n+\cdots+a_1x+a_0$ with $|a_i|\le b, Pr[a_i\ge 0]=\frac{1}{2}$ and  $Pr[a_i\neq 0] =1-r,$ where $Pr$ means probability.
 \subsection{Results}
\begin{figure}[ht!]
\begin{centering}
\includegraphics[width=5.4in]{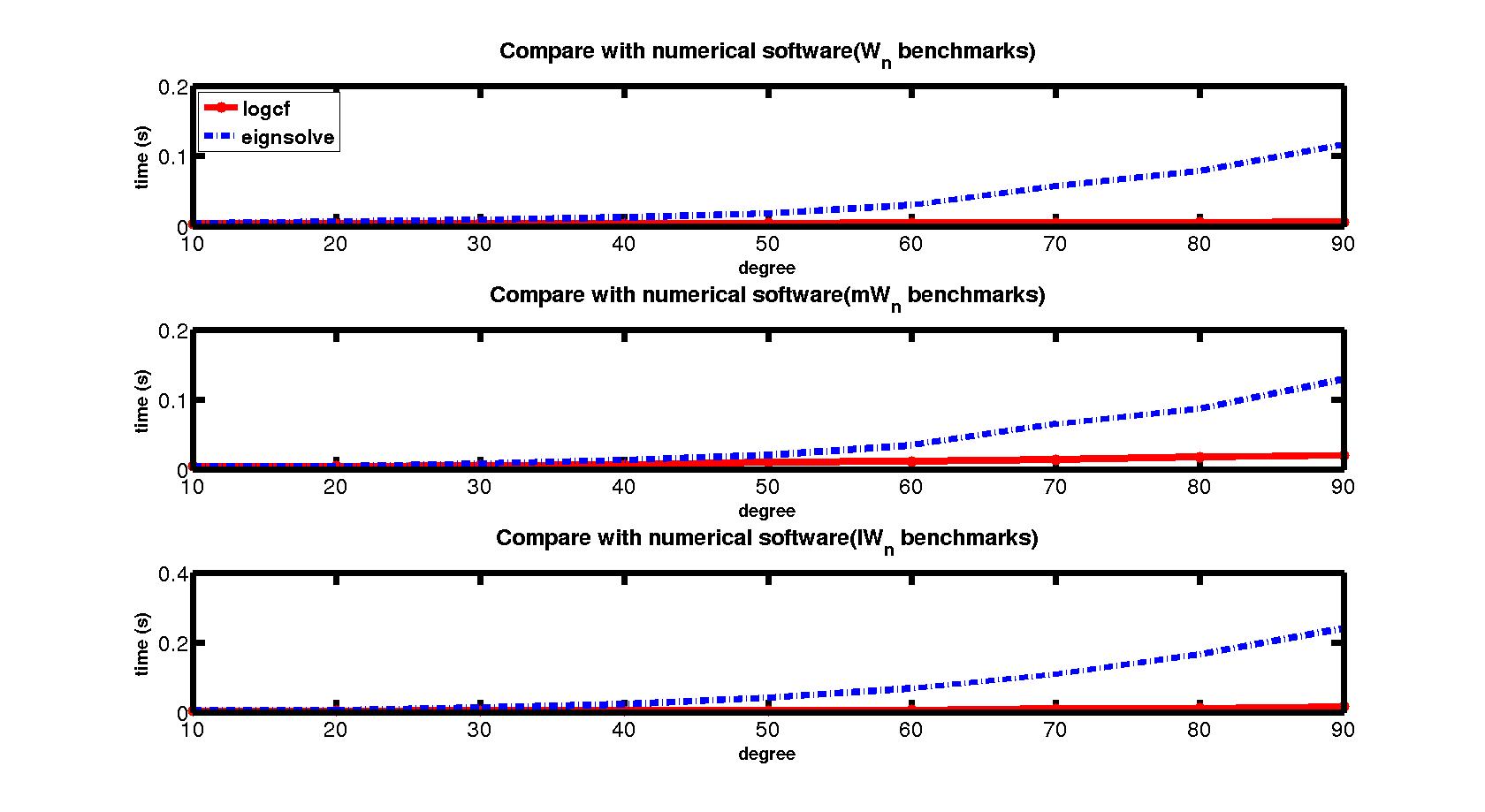}
\caption{ compare with numerical software  \eign \label{fig:1}}
\end{centering}
\end{figure}

\begin{figure}[h!]
\begin{centering}
\includegraphics[width=5.4in]{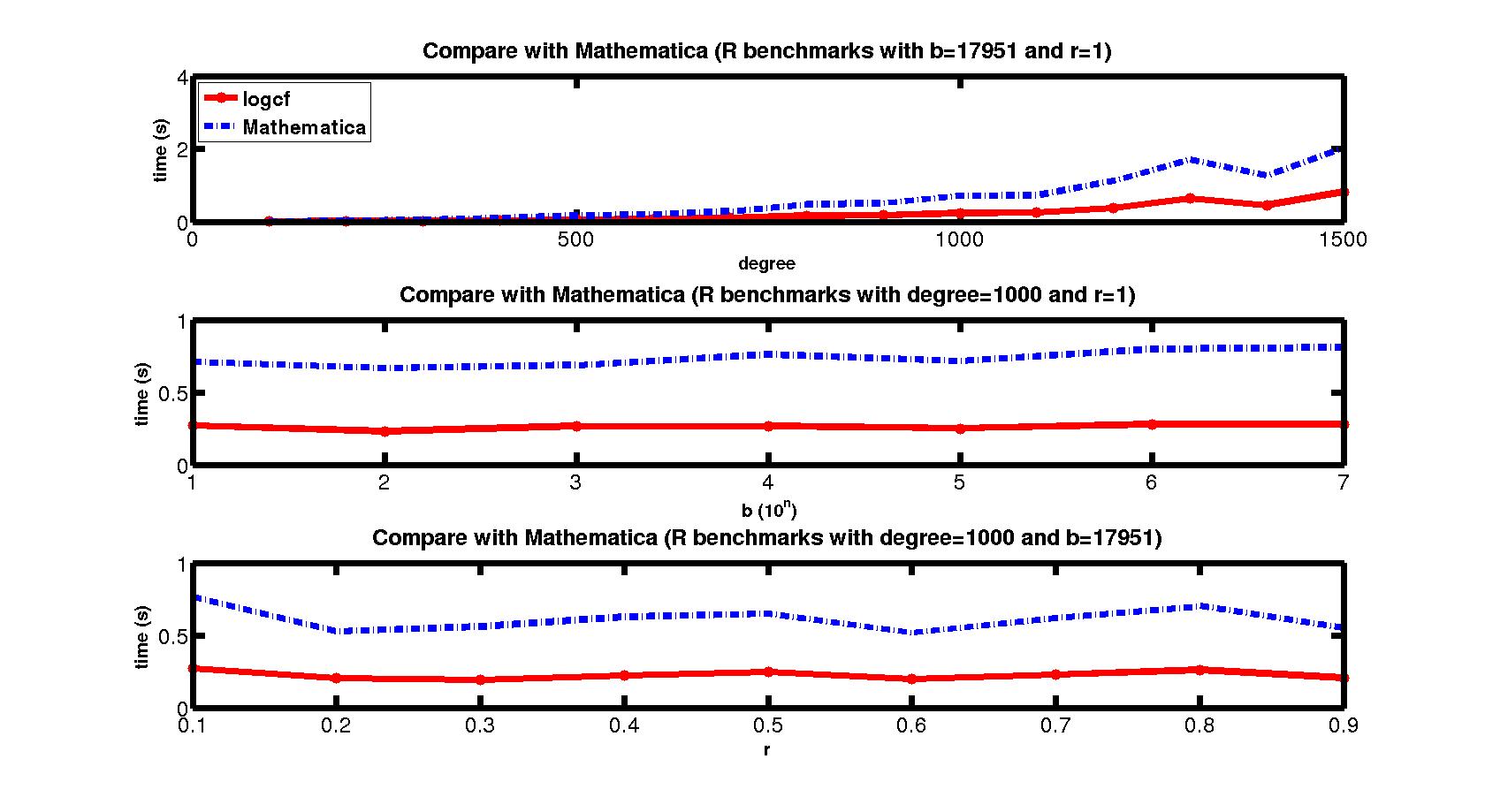}
\caption{ $R(n,b,r)$ benchmarks  with differ setting \label{fig:2}}
\end{centering}
\end{figure}

 The root isolation timings in Tables \ref{tab:mm1}, \ref{tab:mm2} and \ref{tab:open} are in seconds.  Most of the benchmarks we chose have large degrees and the timings show that our tool is very efficient.
As a  built-in \MM\ symbol, \inte\ is    compared with  our tool \froot. The  \MM\  we use has a version number 8.0.4.0. For  almost all
 benchmarks, our  software \froot\  can be  two or three times faster than \inte. The comparative data can be found in Table \ref{tab:mm1}, Table  \ref{tab:mm2} and Figure \ref{fig:2}.
 We also consider open software,  such as \cf\  \cite{hemmer09}, which
 seems to be one of the fastest  open software  available for exact real root isolation. Many experiments  about  state of the art open software for isolating
 real roots have been done in \cite{hemmer09},  which  indicate that     \cf\  is the fastest in many cases.
 In our experiments, \froot\ is much faster than \cf. 
 The comparative result can be found in
 Table \ref{tab:open}. We also compare \froot\  with numerical methods  \eign\ \cite{eigsolev} and \sle\ \cite{hemmer09}. As \eign\ computes all the complex roots, we choose $W_n$, $mW_n$ and $IW_n$ as benchmarks with degrees ranging from 10 to 90, which have only real roots. \sle\ computes only real roots but it has weak stability. Its output on $W_{30}$ only has eight real roots, which is obviously wrong. \sle's running time\footnote{When  running time is very short we run every case for more than ten times and compute the mean.} on $W_{10}$ is $0.022$ seconds and
 $0.024$ seconds on $W_{20}$. In these two cases our software is about $7$ times faster than \sle. We compare \froot\ with  \eign\ and the results are  shown in Figure \ref{fig:1}.
 At the beginning when degree is $10$, the time costs of \froot\ and \eign\ are
 almost equal. As degree becoming larger, the growth rate of our tool's consuming-time is much less than that of  \eign.  When degree reaches $90$, \froot\ is about $20$ times faster than \eign.

 \begin{table}[H]
  \centering
  \begin{tabular}{|| c| c| c|| c|c| c||}
\hline
\scriptsize{Benchmark}  & \scriptsize{\inte}  &\scriptsize{ \froot} &\scriptsize{Benchmark}  & \scriptsize{\inte}  &\scriptsize{ \froot}\\
\hline
$W_ {100}$ & 0.024 & 0.01 & $ IW_{100}$ & 0.048 & 0.01\\
\hline
$W_{200}$ & 0.096 & 0.015 & $IW_{200}$ & 0.148 & 0.015\\
\hline
$W_{300}$ & 0.19 & 0.03 &$IW_{300}$ & 0.33 & 0.03\\
\hline
$W_{400}$ & 0.36 & 0.06 & $IW_{400}$ & 0.72 & 0.08\\
\hline
$W_{ 500}$ & 0.624 & 0.11& $IW_{500}$ & 1.2 & 0.13\\

\hline
$W_{ 1000}$ & 3.33 & 0.87& $IW_{1000}$ & 5.53 & 0.86\\

\hline
$W_{2000}$ & 21.58 & 6.88& $IW_{ 2000}$ & 26.08 & 8.28\\
\hline
$mW_{100}$ & 0.084 & 0.025& $mIW_{100}$ & 0.032 & 0.01\\

\hline
$mW_{ 200}$ & 0.55 & 0.16& $mIW_{200}$ & 0.172 & 0.04\\

\hline
$mW_{300}$ & 1.92 & 0.63& $mIW_{300}$ & 0.548 & 0.16\\

\hline
$mW_{400}$ & 4.92 & 1.77 & $mIW_{400}$ & 1.30 & 0.44\\

\hline
$mW_{500}$ & 10.6 & 4.34 & $mIW_{500}$ & 2.73 & 1.01\\

\hline
$mW_{1000}$ & 140.9 & 65.62 &$ mIW_{1000}$ & 32.9 & 15.56\\
\hline
  \end{tabular}
  \caption{compare with \MM(1)}
  \label{tab:mm1}
\end{table}

\begin{table}[H]
  \centering
  \begin{tabular}{||c| c| c|| c| c| c|| }
\hline

\ \scriptsize{Benchmark}\   & \scriptsize{\inte}  &\scriptsize{ \froot} &\ \scriptsize{Benchmark}\   & \scriptsize{\inte}  &\scriptsize{ \froot}\\

\hline

$T_{100}$&  0.056 &  0.01 & $L_{100}$&  0.072 & 0.02  \\

\hline
$T_{200}$&  0.39 &  0.03& $L_{200}$  & 0.60 &0.16 \\

\hline
$T_{300}$  & 1.29 & 0.10& $L_{300}$  & 2.2 &0.69 \\

\hline
$T_{400}$ & 3.39 &  0.22& $L_{400}$ & 5.64 &1.91 \\

\hline
$T_{500}$ & 7.26 & 0.45& $L_{500}$  & 12.24 &4.59 \\

\hline
$T_{1000}$ & 90.8 & 4.96& $L_{1000}$  & 150 &72.3 \\
\hline

$U_{100}$  & 0.048 &  0.01& $M_{2000}$  & 1.22 & 0.19 \\

\hline
$U_{200}$ & 0.35 &  0.03& $M_{2001}$ & 1.22 & 0.20 \\

\hline
$U_{300}$ & 1.31 & 0.09& $M_{4000}$ & 8.02 & 1.79 \\

\hline
$U_{400}$ & 3.35 &  0.21& $M_{4001}$ & 7.98 & 1.99 \\

\hline
$U_{500}$ & 6.95 & 0.44& $M_{6000}$ & 33.4 & 7.73 \\

\hline
$U_{1000}$ & 87.5 & 4.81&  $M_{6001}$ & 33.7 & 7.82 \\
\hline
  \end{tabular}
  \caption{compare with \MM(2)}
  \label{tab:mm2}
\end{table}


For randomly generated polynomials, we consider different settings of
$(n,b,r)$ as shown in Figure \ref{fig:2}. For each setting $(n,b,r)$, we generate randomly five instances and  compute the mean of five running times. In almost every randomly generated benchmark our \froot\  is two or three times faster than \inte. And We can also  find that
degree is the main factor affecting the  running time.

\begin{table}[H]
  \centering
  \begin{tabular}{|| c| c| c|| c|c| c||}
\hline

\hline
\scriptsize{Benchmark}  &\ \  \ \    \ \ \scriptsize{\cf}\ \  \ \  \ \     & \scriptsize{\froot} &\scriptsize{ Benchmark}  &\ \   \ \ \ \  \ \scriptsize{ \cf} \  \  \ \ \ \    \    & \scriptsize{\froot}\\
\hline
$W_{100}$ & 0.054 & 0.01 &  $IW_{100}$ & 0.056 & 0.01\\
\hline
$W_{200}$ & 0.23 & 0.015 & $IW_{200}$ & 0.20 & 0.015\\
\hline
$mW_{100}$ & 0.054 & 0.025& $mIW_{100}$ & 0.14 & 0.01\\
\hline
$mW_{200}$ & 40.5 & 0.16& $mIW_{200}$ & 2.7 & 0.04\\
\hline
$T_{100}$ & 0.52 &  0.01 & $L_{100}$ & 0.80 & 0.02  \\

\hline
$T_{200}$ & 4.32 &  0.13& $L_{200}$ & 7.50 &0.16 \\

\hline
$U_{100}$ & 0.52 &  0.01& $M_{1000}$ & 43.52 & 0.03 \\

\hline
$U_{200}$ & 4.15 &  0.12& $M_{1200}$ & 88 & 0.05 \\

\hline

\hline
  \end{tabular}
  \caption{compare with \cf}
  \label{tab:open}
\end{table}

\section*{Acknowledgements}
The work is partly supported by NSFC-11271034,  the ANR-NSFC project EXACTA (ANR-09-BLAN-0371-01/60911130369)
and the project SYSKF1207 from ISCAS.
The authors would like to thank Steven Fortune who sent us the source code of \eign\ and Elias P. Tsigaridas who helped us compile \cf.

\end{document}